\documentclass[12pt,preprint]{aastex}

\newcommand{\eg}{{\it e.g.,}}
\newcommand{\ie}{{\it i.e.,}}

\begin{document}

\title{Adaptive Optics Imaging and Spectroscopy of Cygnus A: I.\ Evidence for a Minor Merger\footnotemark[1]}

\footnotetext[1]{Based on observations with the NASA/ESA Hubble Space 
Telescope,
obtained from the data archive at the Space Telescope Science Institute.
STScI is operated by the Association of Universities for Research in Astronomy,
Inc. under NASA contract No. NAS 5-26555.}

\author {Gabriela Canalizo\altaffilmark{2}, Claire Max\altaffilmark{2,3},
David Whysong\altaffilmark{4}, Robert Antonucci\altaffilmark{4}, 
Scott E. Dahm\altaffilmark{5}}

\altaffiltext{2}{Institute of Geophysics and Planetary Physics, Lawrence 
Livermore National Laboratory, 7000 East Avenue, L413, Livermore, CA 94550}

\altaffiltext{3}{Center for Adaptive Optics, University of California, Santa 
Cruz, Santa Cruz, CA 95060}

\altaffiltext{4}{Physics Department, University of California, Santa Barbara, CA 93106}

\altaffiltext{5}{Institute for Astronomy, University of Hawaii, 2680 Woodlawn
 Drive, Honolulu, HI 96822}

\begin{abstract}
We present Keck II adaptive optics near infrared imaging and spectroscopic 
observations of the central regions of the powerful radio galaxy Cygnus A.  
The $0\farcs 05$ resolution images clearly show an unresolved nucleus 
between two spectacular ionization/scattering cones.
We report the discovery of a relatively bright ($K'\sim19$) secondary
point source 0\farcs4 or 400 pc in projection southwest of the radio nucleus. 
The object is also visible
in archival {\it Hubble Space Telescope} optical images, although it is
easily confused with the underlying structure of the host.   Although the 
near infrared colors of this secondary point source are roughly consistent 
with those of an L-dwarf, its spectrum and optical-to-infrared spectral energy
distribution (SED) virtually rule out the possibility that it may be any 
foreground
projected object.   We conclude that the secondary point source is likely to 
be an extragalactic object associated with Cygnus A.   We consider several
interpretations of the nature of this object, including: a young star cluster 
peering through the dust at the edge of one of the ionization cones;  an 
older, large globular cluster; a compact cloud of dust or electrons that is 
acting as a mirror of the hidden active nucleus; and the dense core of a gas 
stripped satellite galaxy that is merging with the giant elliptical host.
The data presented here are most consistent
with the minor merger scenario.  
The spectra and SED of the object suggest that it may be a densely packed 
conglomeration of older stars heavily extincted by dust, and its
high luminosity and compact nature are consistent with those of a 
satellite that has been stripped to its tidal radius.
Further spectroscopic observations are nevertheless necessary to confirm 
this possibility.   
\end{abstract}


\section{Introduction}

Mergers and strong interactions of galaxies are often cited as important
triggering mechanisms for activity in radio galaxies.   Many powerful radio
galaxies exhibit dramatic signs of interactions 
\citep[\eg\ ][]{hec86,smi89,rid97}, but many others appear to reside in
early type galaxies that show no signs of interaction \citep[\eg\ ][]{dun03}.
One interesting alternative is that, at least in some active nuclei, the 
fueling may be driven by minor mergers \citep[][and references therein]{tan99}.
Signs of tidal interaction in these minor mergers would be more elusive, 
and their detection may require deep, high resolution observations 
attainable only with the {\it Hubble Space Telescope (HST)} and, at 
ground-based facilities, with the use of adaptive optics (AO).

The prototype radio galaxy Cygnus A is a Fanaroff-Riley class II radio
galaxy showing two lobes extending to more than 60 kpc
on either side of a giant, seemingly non-interacting (though gas rich and 
star-forming) elliptical host. 
Because of its proximity ($z=0.056$) and extreme characteristics, Cygnus A 
has played a
fundamental role in the study of virtually every aspect of powerful
radio galaxies \citep[see][for a review]{car96}.
Different lines of evidence indicate that Cygnus A harbors a 
heavily extincted quasar \citep[\eg ][]{ogl97,ant94}.  However, 
the giant elliptical shows an r$^{1/4}$ radial surface brightness profile
\citep[\eg ][]{sto94} and no large scale tidal features indicative 
of a recent merger event that might have triggered the nuclear activity.

We have obtained near infrared Keck II AO images of Cygnus A 
of unprecedented 
resolution ($\sim0\farcs 05$ or 50 pc for $H_{0}=70$ km s$^{-1}$ Mpc$^{-1}$,
assumed throughout the paper) and depth which reveal new exciting 
information.   In this paper,
we report on the discovery of a secondary point source in the
central regions of Cygnus A, which may be the smoking gun of a
minor merger event.   In a subsequent paper 
(Canalizo et al., in preparation; hereafter Paper II) we
discuss the spectroscopy of the nucleus, the morphology of the central
regions, and the interpretation of our observations in the context of 
unified models.

\section{Observations and Data Reduction}

Near infrared spectroscopic and imaging observations of the central regions
of Cygnus A were carried out using the Keck II AO system
\citep[][]{wiz00a,wiz00b,joh00}
with the NIRC-2 camera (PI: K. Matthews \& T. Soifer).
The Keck II AO system is located on an optical bench at the 
Nasmyth platform of the 10-m telescope.  The Xinetics deformable 
mirror has 349 degrees of freedom, of which approximately 249 are 
illuminated at any given time as the hexagonal pupil of the telescope 
rotates on the round deformable mirror.  The Shack-Hartmann wavefront 
sensor is based on a 64$\times$64 pixel Lincoln Laboratories CCD with read 
noise of approximately 6 electrons per pixel.  The real-time computer 
is based on the Mercury RACE architecture, and uses sixteen Intel 
i860 floating-point CPUs.  The control system's closed-loop bandwidth 
is typically about 30 Hz using natural guide stars ($V\lesssim 13.5$) as 
a wavefront reference.  For the spectroscopic and imaging observations 
of Cygnus A, we guided on a $V\sim13$ magnitude star $29\farcs4$
southwest of the nucleus.   The sampling frequency of the wavefront 
sensor was 55 Hz.

Cygnus A was imaged in $K'$ on UT 26 May 2002 through the NIRC-2 narrow
field camera,
which yields a plate scale of $0\farcs01$ pixel$^{-1}$.  The total integration
time was 5$\times$300~s, and the FWHM of stars near Cygnus A was $0\farcs05$.
Imaging in $J$- and $H$-bands was done on UT 27 May 
2002 through the wide field camera, yielding a plate scale of $0\farcs04$ 
pixel$^{-1}$; total integration times were 6$\times$60~s in $J$ and 
10$\times$60~s in $H$.   The FWHM in these images ranged from $0\farcs07$
near the guide star (GS) to $0\farcs1$ near Cygnus A in $H$-band, and 
from $0\farcs09$ to $0\farcs14$ in $J$-band
All the observations were obtained under photometric conditions.
Standard stars from \citet{per98} were observed throughout each night.

Spectroscopic observations in $K'$ and $J$ bands were obtained with NIRC-2
on UT 4 and 5 August 2002 respectively, using the wide field camera 
($0\farcs04$ pixel$^{-1}$) and
the low resolution (LOWRES) grism.
The width of the slit was 0\farcs08, projecting to 2 pixels and yielding
a resolution of 32 \AA\ per resolution element (2 pix) for the $K'$ spectrum, 
and 20 \AA\ per resolution element for the $J$ spectrum
(R$\sim$700 in each case).
Total integration times were
2$\times$1200~s for $K'$ and 2$\times$900~s for $J$.
The choice of slit position angle was constrained by a number of limitations.
Since the GS was nearly 30\arcsec\ away from Cygnus A, 
close to the limit for
the steering mirrors, Cygnus A could only be positioned 
at certain places on the detector, and at certain angles that would maximize
the possible distance between the steering mirrors and the GS.  We chose 
a PA that would allow us to place the object in a region of the detector
that would maximize the wavelength coverage, but that would at the same time 
minimize the flux contamination from the surrounding emission.  The resulting 
slit PA was 140$\arcdeg$.  Nearby B8 and A0 stars were observed 
immediately after each 
science observation and used for telluric correction and flux calibration.   
These stars were also imaged through $J$ and $K'$
filters and calibrated against photometric standards from \citet{per98}.

The spectra were reduced with IRAF, using standard reduction procedures.
After correcting for dark current and bad pixels, flat fielding, and
subtracting sky, we calibrated the 2-d spectrum in wavelength using
the OH sky lines.  We rectified the spectra by tracing the spectrum
of a bright standard placed at the same position along the slit as
the science target.   The extracted science spectra were then corrected
for atmospheric absorption using telluric standards.

\section{A Secondary Point Source in the Core of Cygnus A\label{merging}}

\begin{figure}[htp]
\epsscale{0.8}
\plotone{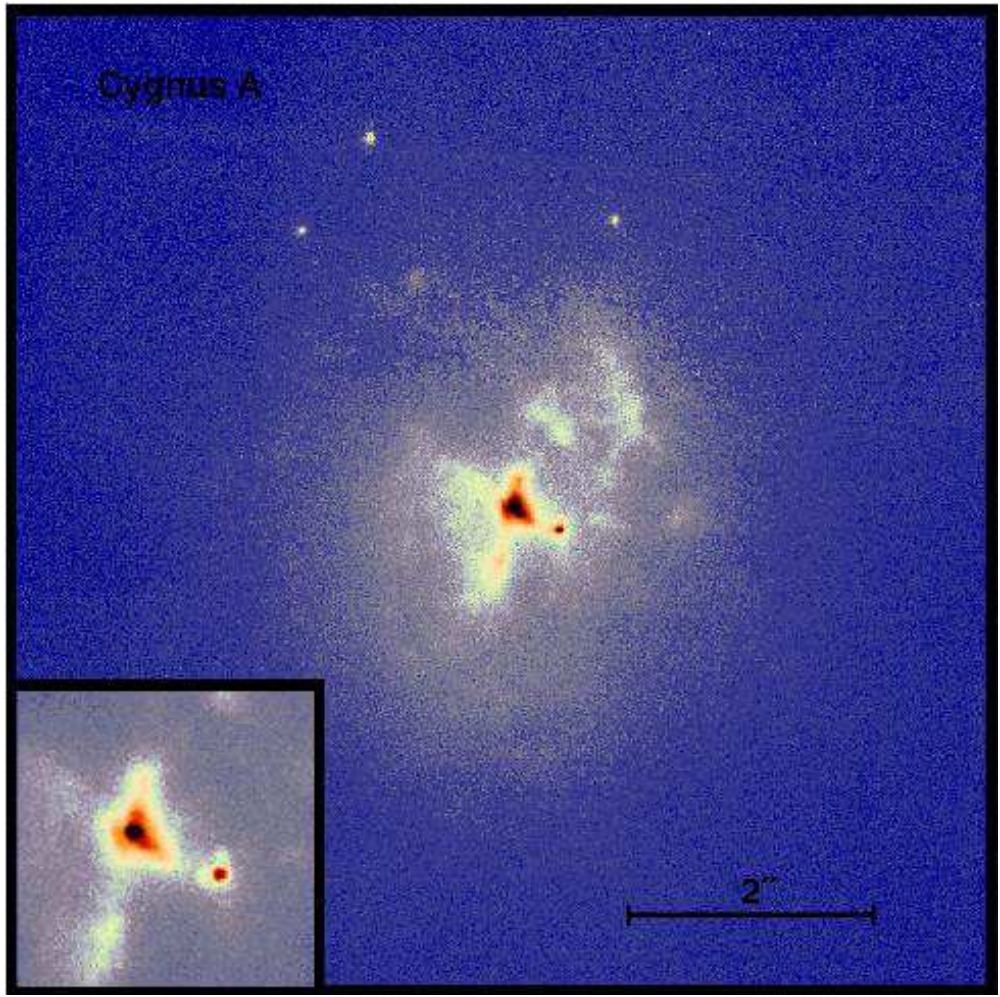}
\caption {False color NIRC-2 AO $K'$ image of the core of Cygnus A. In
this and the following images North is up and East is to the left.  The
inset is 1\farcs3 $\times$ 1\farcs3 and shows the central region at a
lower contrast to show the point-like nucleus as well as a secondary 
point source 0\farcs4 to the west-southwest.  
\label{knarrow}}
\end{figure}

Figure~\ref{knarrow} shows a NIRC-2 AO $K'$ image of the central 
$8\arcsec\times8\arcsec$ region of Cygnus A.   The unresolved nucleus
is clearly seen between two spectacular ionization/scattering cones.
As noted above, we postpone our analysis of these and other features 
in the central regions of Cygnus A to Paper II. However, on the suggestion 
of the referee, we provide a few remarks on the cones in Section~\ref{cones}.

In addition to the ionization cones, 
we find a bright secondary point source $\sim 0\farcs 4$ or $\sim 400$ pc
(in projection) 
west-southwest of the nucleus.   What is the nature of this object?   While 
the image is suggestive of a double nucleus in Cygnus A, we must consider all 
possibilities.    First, it is conceivable that this object may be a 
faint foreground star in our Galaxy, particularly since Cygnus A is near 
the plane of the Galaxy ($b=5.76\arcdeg$).

\subsection{Secondary Point Source: a Projected Galactic Object?}

The first step in investigating whether the secondary point source is
a foreground object is to obtain its photometric colors and compare them
to those of different galactic objects.   Obtaining accurate photometry
in AO images is challenging because of the difficulty in determining
the point spread function (PSF).
We attempted to obtain a
suitable PSF that would account for anisoplanatism by observing pairs of
stars where the GS had the same magnitude, position angle, and separation
from the PSF as the GS used for Cygnus A.   However,
both the atmosphere and the performance of the AO system change rapidly
between exposures, so it was not possible to obtain a good PSF using
this method.
Ideally, using a large aperture including most of the halo of the PSF 
would yield the most accurate photometry in AO imaging. 
It is obviously impossible to do this in the case of 
the secondary point source, since the object has a bright and highly 
structured background. 

In order to circumvent these difficulties, we used as PSFs field stars from
each image of Cygnus A.
We chose two stars at the same PA from the GS as the core
of Cygnus A, one closer and the other farther away, to account for the 
effects of anisoplanatism.   The stars were only a few arcseconds from
Cygnus A and there was
some contamination from the outskirts of the galaxy.   We subtracted
this contamination by fitting a low order polynomial to the background.   
We then calibrated these stars against standard stars by using a large 
(80 pixels) aperture.   Next, we
scaled the stars to match the secondary point source, and subtracted
them.   Finally, we used the scaling factor to obtain the flux
of the secondary point source.
We obtained consistent results using either PSF star.   
In order to further test the consistency of the method, we measured 
photometry of other stars in the field in this way and compared
it with large aperture photometry;
in every case, we obtained magnitudes consistent within 0.02 mag,
even in the $J$ image, where anisoplanatism is more significant.  

The main source of uncertainty was determining the
best PSF subtraction, since faint PSF residuals could be easily
confused with the highly structured background underlying the secondary
point source.   In every case we achieved good fits
to the point source, apparently leaving no residuals.  However, we
chose to estimate uncertainties by
considering scaling factors that lead to obvious over- and 
undersubtractions of PSFs.   The errors we quote as 1$\sigma$ should 
then be regarded as very conservative.

We used the same method to measure photometry of the nucleus,
obtaining larger errors since the background is more confused by
the bright underlying structure.
We list in Table~\ref{phototab} the photometry of the
nucleus and that of the secondary point source, as well as
their colors transformed to the CIT system \citep[][]{per98}.   
These magnitudes have not been corrected for Galactic extinction 
\citep[A$_V$=1.263 towards Cygnus A;][]{sch98}, since we are 
making no assumptions as yet about the actual position or nature 
of the object.   The $K'$ flux we obtain for the nucleus is nearly
a factor of two lower than the 2.25 $\mu$m flux measured by 
\citet{tad99}, and nearly 10 times lower than the $K-$ band flux
measured by \citet{djo91}.  This is not surprising since,
due to the lower resolution in the $HST$ and ground-based images,
both of
these measurements are expected to have more contamination from
the bright structure underlying the nucleus.   If, for example,
we measure the flux of the nucleus in our $K'$ AO images using a 
0\farcs1 aperture, we
obtain a flux nearly identical to that measured by \citet{tad99} 
in the $HST$ images.   This point, and the fact that the nucleus is well
fitted by a point source, will be discussed in more detail in Paper II.

\clearpage

\begin{deluxetable}{lcc}
\tablewidth{3.5in}
\tablecaption{Photometry of Cyg A Nucleus and Secondary Point Source\label{phototab}}
\tablehead{\colhead{} & \colhead{2nd Pt. Source} & \colhead{Nucleus}}
\startdata
$V$                                & 25.03$\pm$0.31 & \nodata        \\
$m_{F622W}$                        & 25.42$\pm$0.17 & \nodata        \\
$I_{\rm C}$                        & 23.41$\pm$0.14 & \nodata        \\
$J$                                & 20.61$\pm$0.11 & 21.15$\pm$0.21 \\
$H$                                & 19.98$\pm$0.12 & 20.42$\pm$0.15 \\
$K'$                               & 19.14$\pm$0.08 & 18.50$\pm$0.14 \\
($J-H$)$_{\rm CIT}$\tablenotemark{a} &\phn0.62$\pm$0.23 &\phn0.71$\pm$0.35 \\
($H-K$)$_{\rm CIT}$\tablenotemark{a} &\phn0.82$\pm$0.21 &\phn1.87$\pm$0.29 \\
\enddata
\tablenotetext{a}{Transformation to the CIT system using equations given
by \citet{per98} and assuming $K = K'$}
\end{deluxetable}

\clearpage

\begin{figure}[htb]
\epsscale{0.7}
\plotone{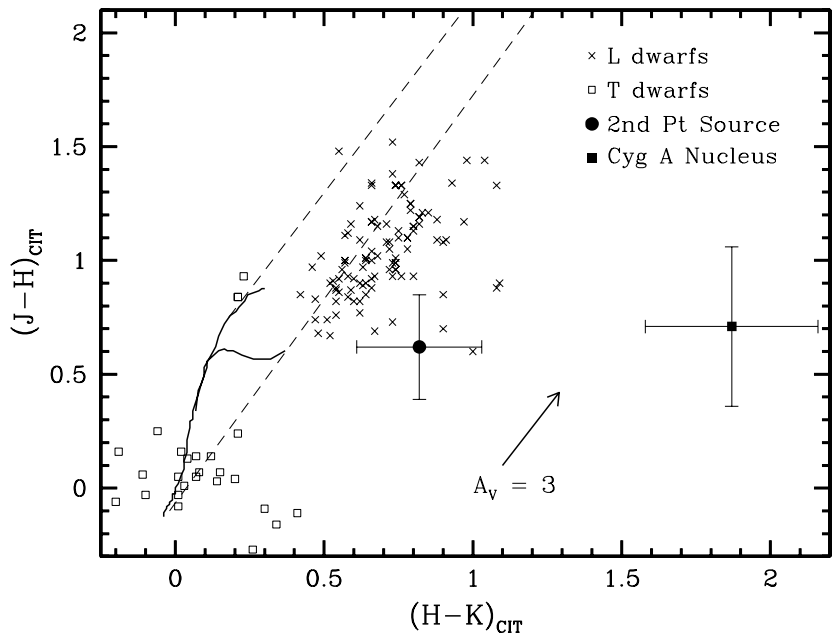}
\caption {Color-color near infrared diagram
in the CIT photometric system.  The solid traces mark
the main sequence (O9 through M6; lower branch) and giants
(G0 through M7 upper branch).   Dashed lines indicate the direction
of galactic reddening, and a galactic reddening of A$_V$=3
is indicated by the arrow.   L-dwarfs are plotted as crosses, and
T-dwarfs as open squares.   
The secondary point source is plotted as a solid circle, and the
nucleus of Cygnus A is plotted as a solid square.   
\label{2color}}
\end{figure}

In Fig.~\ref{2color} we show the position of the secondary point source
in the near infrared (NIR) color--color diagram.   Also shown are the main
sequence, dwarf and giant branches as given by \citet{tok00}.  Very low 
mass objects are also plotted: L-dwarfs from \citet{dah02} and \citet{kir00}, 
and T-dwarfs from \citet{dah02} and \citet{bur02}.
Figure~\ref{2color} shows that the NIR 
colors of the secondary point source can only be consistent with a
somewhat unusual L-dwarf if the object is a foreground star.   
If any amount of Galactic reddening is taken into account, the object
is moved even farther away from the L-dwarf locus.   Pre-main sequence (PMS) 
stars can also occupy this region of the diagram.   However, PMS objects
are always associated with star forming regions, and there are no such 
regions near the position of Cygnus A.    Data from Two Micron All Sky Survey
(2MASS) shows that a relatively small number of faint galaxies also fall 
in this region \citep[see, \eg ][]{jar00};
we will return to this possibility in section~\ref{extragalactic} when we
consider different extragalactic objects.

An unabsorbed L-dwarf with 
($J-K$)$_{CIT}$ = 1.42 would be expected to have an absolute magnitude $M_{J}$ 
roughly between 11 and 13.5 \citep[][]{dah02}, which in turn would
indicate that, if this object were an L-dwarf, it would be located
at a distance of 200--800 pc.   The object is visible in $HST$ WFPC2
images (see below) and the first of these
images was taken almost six years prior to our NIRC-2 image.   We
measure no displacement in six years.   The proper
motion of this object is then less than 6.6$\times10^{-3}$ \arcsec yr$^{-1}$,
implying a tangential velocity of less than 6 km s$^{-1}$ if the
object were at 200 pc, or 25 km s$^{-1}$ if it were at 800 pc.  
Although the implied velocity is somewhat lower than those expected
for disk stars and those found for other L-dwarfs 
\citep[\eg $<V_{tan}> = 31 \pm 4$ km s$^{-1}$ in ][]{dah02}
we cannot, on this basis alone, rule out the possibility that the object 
may indeed be an L-dwarf.

\begin{figure}[htp]
\epsscale{1.0}
\plotone{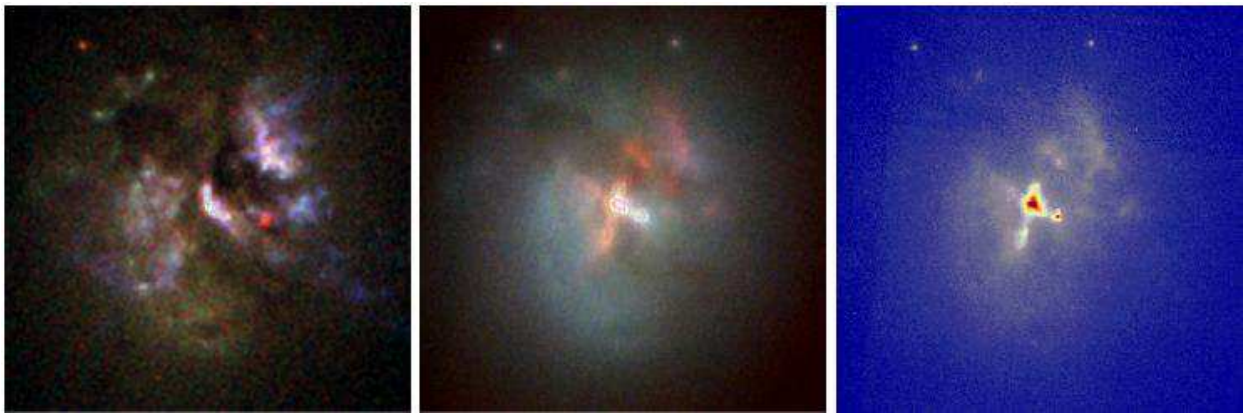}
\caption {Three-color WFPC2 PC1 (left panel) and Keck NIRC-2 (central panel) 
images of the core of Cygnus A.  For the WFPC2 PC1 image, red is the F814W
filter, green is F622W, and blue is F555W.  For the Keck NIRC-2 image,
red is $K'$, green is $H$, and blue is $J$. 
A $K'$ image is shown (in false color, right panel) for reference. Each image 
is 5\farcs 9 $\times$ 5\farcs 9.
\label{colorcyga}}
\end{figure}

Through careful astrometry, we were able to identify the secondary 
point source in archival $HST$ WFPC2 images.   Figure~\ref{colorcyga}
shows a three-color composite optical image of Cygnus A (left panel), 
where red corresponds to F814W, green to F622W, and blue to F555W 
\citep[individual images first appeared in][]{jac98,jac96}.  
We also show in the 
central panel of Fig.~\ref{colorcyga} a three-color near infrared image from 
our NIRC-2 data, where red is $K'$, green is $H$, and blue is $J$.  The 
narrow field camera $K'$ image is included on the right panel at the same
scale for reference.  

\begin{figure}[ht]
\epsscale{0.7}
\plotone{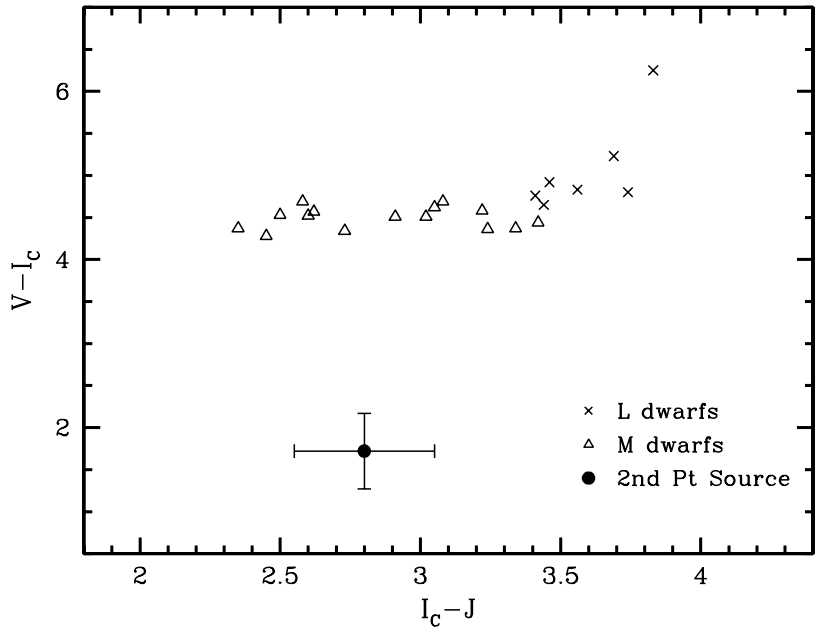}
\caption {Color-color diagram in the CIT and Cousins photometric systems.  
L-dwarfs are plotted as crosses, and M-dwarfs as open triangles.   The
1$\sigma$ errors in the colors for these objects are typically smaller 
than the size of their symbols.
The secondary point source is plotted as a solid circle, showing that
the SED of this object is too blue to be an L-dwarf. \label{plotvij}}
\end{figure}

We have used the position
measured from our narrow field camera $K'$ AO image (Fig.~\ref{knarrow})
to obtain optical photometry by fitting a PSF to the secondary point 
source in the WFPC2 images.  
For consistency, we used the same fitting method and PSF stars that
we used for the NIR AO data.  The resulting
photometry is listed in Table~\ref{phototab}.   The optical spectral
energy distribution (SED) of this object is bluer than those of
L-dwarfs.   In Fig.~\ref{plotvij} we show an $I_C - J$ vs. $V - I_C$
diagram where we plot M- and L-dwarfs from \citet{dah02} (T-dwarfs are
off this diagram since their $V - I$ values are much larger).
The secondary point source is clearly set apart at shorter wavelengths
from these Galactic low mass objects.   Therefore, based on its
photometry, we conclude that it is highly unlikely that the secondary 
point source in Cygnus A is a foreground Galactic object.

The reader will notice that the composite color {\it HST} image in 
Fig.~\ref{colorcyga} shows a very
red object west of the secondary point source.   This object appears
so red because it is not 
visible in either the F622W or the F555W images.
The F814W image was taken on UT 1996 April 8, roughly 2 and 1.5 years
after the the first two were obtained.   The red object is therefore very
likely a supernova (SN) that went off some time after the first two sets of
images were taken. We measure an absolute I magnitude  
M$_{I} = -15.43 \pm 0.09$ (correcting for Galactic reddening, 
but neglecting the $k$ correction, which should be smaller than 0.1 magnitude).
This magnitude is typical of a Type I or II SN that is roughly
between 50 and 120 days old \citep[][]{whe00}.   The next set of {\it HST}
observations of Cygnus A was obtained with NICMOS on UT 1997 December 15 
\citep[][]{tad99,tad00}, when the SN was $\sim 700$ days old and would be 
at least
8 magnitudes fainter, which is why it is not visible in the 1997 NICMOS 
images.   It is not possible to determine whether this is a type I or II
SN based on this image alone.   Type Ia SNe are believed to be the only 
type of SN occurring in
quiescent elliptical galaxies \citep{van91}, presumably because older
stellar populations dominate these galaxies.  They may also be
common near (and possibly triggered by) radio jets \citep{cap02}.
However, the host galaxy of Cygnus A is known to have regions
of star formation \citep[\eg ][]{fos99,jac98, sto94}, so this
SN could potentially be a type II SN.

\subsection{Secondary Point Source as an Extragalactic Object
\label{extragalactic}} 
If the secondary point source is not a projected Galactic object,
then it is likely to be an extragalactic object associated with Cygnus A.
A background galaxy would be virtually undetectable in the highly
extincted central regions of Cygnus A, unless the extinction is extremely 
patchy.
The bright and compact nature of the secondary point source might suggest 
that it may be a massive young star cluster in the host galaxy of
Cygnus A.  It is conceivable that many such clusters could be
present in the circumnuclear regions of Cygnus A, deeply obscured by
dust.  In this scenario, the secondary point source would be visible
due to its strategic location near the edge of the cone, where the
extinction might be less severe.

\begin{figure}[ht]
\epsscale{0.7}
\plotone{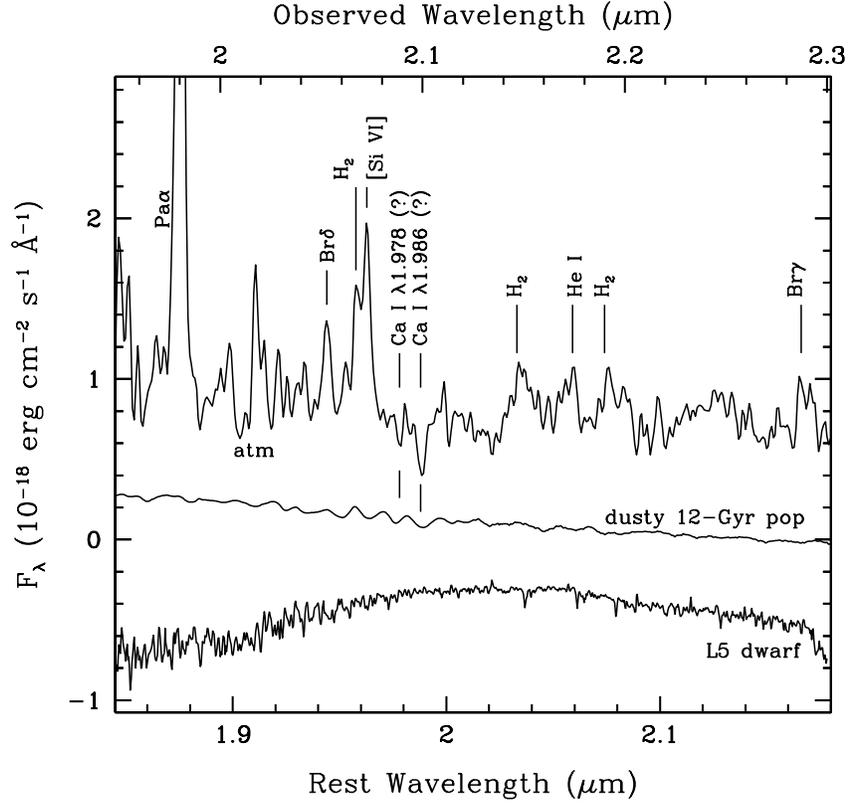}
\caption {$K'$ spectrum of secondary point source in Cygnus A (top trace).
The top axis indicates the observed wavelength, and the bottom axis
the rest frame wavelength, assuming $z=0.056$.   
The spectrum has been smoothed using a gaussian of $\sigma=1$ pixel.
The strong emission lines come from the surrounding gas.   The location
of two possible absorption features is indicated.  The middle trace
is a reddened GISSEL00 model \citep[][]{bru03} of a 12-Gyr old population at 
the redshift of Cygnus A (see text for details).
The bottom trace is an L5 dwarf from \citet{mcle01} plotted with respect to the top (observed wavelength) axis.
The 12-Gyr old stellar population and the L5 dwarf spectra are normalized 
to match the flux of the secondary point
source, and are shifted downward to allow for comparison.\label{kcomp}}
\end{figure}

The $K'$ spectrum (Fig.~\ref{kcomp}) shows a $\sim 5\sigma$ (per resolution
element) continuum at the 
position of the secondary point source, with a spatial FWHM of 
$\sim0\farcs 08$, roughly consistent with the FWHM of $\sim0\farcs 07$
measured in the wide camera $K'$ image obtained the same night.   
Similarly, the $J$-band spectrum shows a $\sim 3\sigma$ continuum.  
It is difficult to determine whether there are emission lines intrinsic
to the secondary point source, as there is much extended emission 
associated with the ionization cone around the object (Paper II).   The 
$K'$ spectrum 
shows strong emission lines, identified in Fig.~\ref{kcomp}; however, 
the (spatial) peak of the emission
is centered 1.5 pix or 0\farcs06 southeast of the continuum, and therefore
appears to come from the edge of the ionization cone, which underlies
the continuum object.  The same is true for the $J-$band spectrum, which
shows strong [Fe II] emission at 1.258 $\mu$m.  
 Therefore we cannot use these emission lines to measure a
redshift for the secondary point source, although we also cannot rule
out the possibility that the object may indeed have fainter emission
lines at the same redshift.
A young star cluster, however, would be expected to be embedded in 
gas.  This gas would be ionized by the hot stars forming an H\,II
region, and the spectrum would show strong emission lines associated
with the continuum object.    

\clearpage
\clearpage

The conspicuous absence of emitting gas in this object suggests that
it may instead be a conglomeration of older stars.
The signal to noise ratio of the spectrum of this 
$K'$=19.1 object is not high enough to allow us to 
identify unequivocally stellar absorption lines. In Fig.~\ref{kcomp} 
we have plotted an isochrone synthesis model from the 
GISSEL00 empirical models \citep[][]{bru03}.  The model is a 12-Gyr old 
population at the redshift of Cygnus A ($z\sim0.056$).   We have also 
simulated the presence of
embedded dust (which is obviously present in Cygnus A) by reddening
the spectrum using the ``dusty galaxy'' models calculated by \citet{wit92}.
In these models, the dust and the stars are both assumed to have constant
density within a sphere, and the only variable is the optical depth to
the center at a specific wavelength; in this case we choose a $V$-band
central optical depth of 6.  While these models are highly artificial,
they show the general nature of the reddening due to embedded dust.  
The final spectrum shows very roughly the same continuum shape as that of
the secondary point source in Cygnus A, and two of the absorption features
in the \ion{Ca}{1} triplet (those at 1.978 $\mu$m and 1.986 $\mu$m) are
coincident with potential absorption features
in the spectrum of the secondary point source
(the third line, if present, is swamped by the extended [\ion{Si}{6}]
emission).  The 12-Gyr model is also a fair fit to the overall SED of the 
object (from $V$ to $K$ bands), except at $J$-band, where the flux of the
model falls 2$\sigma$ below the photometric point.
Thus the spectrum could be consistent
with a reddened older stellar population at the redshift of Cygnus A,
but we will need deeper spectra to confirm the nature of this object.

A bright compact object formed by older stars could be a large globular 
cluster, although it seems unusual that we would not observe any other 
such clusters throughout the host galaxy.  
It is difficult to estimate the luminosity of the 
secondary point source since we do not 
know how much extinction there is at this point.   But even without
taking into account any extinction in Cygnus A, the secondary point
source has an absolute $K$ magnitude of M$_K = -17.92$.  This is
almost 500 times more luminous than the most luminous globular cluster
in our Galaxy, $\omega$ Centauri \citep[NGC\,5139][]{har96},
which has an M$_K = -10.22$, 
and $\sim4000$ times more luminous
than the average globular cluster in our Galaxy.   Therefore, if the
secondary point source in Cygnus A is indeed a conglomeration of 
older stars, it must be significantly more massive than a typical
globular cluster.  However, given its apparent size ($\lesssim 50$ pc),
it must have similar or higher densities than a globular cluster.

Such conditions (\ie\ a large, densely packed group of older stars)
can be found in the cores of low luminosity galaxies.    At least 
in the case of ellipticals, low luminosity galaxies have much denser
central regions
than their high luminosity counterparts \citep[see, \eg ][]{fab97}.   
For example, 
a galaxy initially having one tenth the mass of the Cygnus A host
would have a space density 20 times higher than that of the host at a radius
of 1 pc, while a galaxy with a 1:100 mass ratio would have a 1:51 density 
ratio at the same radius \citep[][]{hol99}.
\citet{ken03} investigate the dynamics of minor mergers for active galactic 
nuclei and estimate the mass loss of an infalling satellite simulated
by a Hernquist sphere (see their equation 6).
This estimate is simply obtained 
by equating the mean density of the satellite within
its tidal radius to the
mean density of the galaxy within the radius given by the distance between
the satellite and the center of the primary galaxy. 
In the simulation, the satellite is truncated to its tidal radius
and the density structure internal to this point remains unchanged.
Although this assumption is unphysical, \citet{ken03} argue that it
is a reasonable approximation since the satellite will react to the
density truncation on an internal dynamical timescale which is of the 
order of the orbital timescale.
Following their method, we estimate the mass loss due to tidal stripping 
for satellites of different mass
ratios with respect to Cygnus A.   We simulate the Cygnus A host galaxy
using the S\'{e}rsic density profile as given by \citet{lim99} in their
equation 17, with $\nu = 0.25$ and p = 0.851, which correspond to a
de Vaucouleurs profile.   We use the galaxy parameters for Cygnus A 
given by \citet{mar03},
so that $M_{bul}=1.6\times10^{12}$ M$_{\sun}$, $R_{eff}= 31$ kpc, and 
$\rho_{o} = 3.09\times10^{3}$ M$_{\sun}$ pc$^{-3}$.  For satellites,
we scale this density profile to match the parameters given by 
\citet{mar03} for NGC\,3115, NGC\,4742, and M\,32,
which have mass ratios of 1:9, 1:145, and 1:1667 respectively, with 
respect to the Cygnus A host.   We find that the 1:9 satellite loses 
$\sim$94\% of its 
mass by the time it reaches a distance of 400 pc to the nucleus;
its remaining mass in the tidally stripped core is
roughly $7\times10^{-3}$ that of the host.  Similarly, the 1:145
satellite loses $\sim$95\% of its mass and the remaining mass ratio 
is $\sim4\times10^{-4}$ with respect to the host.  
Because of its much higher central density, the 1:1667 satellite
only loses $\sim$60\% of its mass, and the resulting mass ratio is
2.4$\times$10$^{-4}$. 
Simulating the satellites with Hernquist spheres scaled to 
match their observed parameters leads to similar results,
within a factor of 2.  

For comparison, we measure an $H$-band absolute magnitude for the Cygnus A
giant elliptical host galaxy of M$_H \sim -25.2$, roughly consistent with
the M$_R = -23.7$ measured by \citet{car89} (transformed to our choice of
Cosmology), if the host has an SED typical of ellipticals
(we use $H$ rather than $K'$ since our wide field $H$-band image is deeper).
Neglecting the 
effects of extinction and assuming a similar 
mass to light ratio in the host and the putative merging galaxy, 
the latter would have roughly 6$\times10^{-4}$ the mass of the host.  
This is comparable to the mass ratios we obtain for the tidally stripped 
1:145 and 1:1667 companions at a
distance of 400 pc from the center of Cygnus A.   
All of these are obviously gross approximations and
oversimplified assumptions, and they should only serve to provide
some reference for the scales that we are discussing, and to show that
the interpretation of the secondary point source as an infalling satellite
is not unreasonable.
Cygnus A is associated with a rich cluster of galaxies \citep[possibly of
Abell richness class 3;][]{owe97}.  Thus a merger between the giant elliptical
host and a galaxy having a mass 100 or 1000 times smaller should not be
an unlikely event.

The secondary point source in Cygnus A may then be the gas stripped 
core of a merging galaxy which has thus far survived a merger event.
\citet{hol00} consider the effects of a central massive black hole
in the merger between an elliptical galaxy and a smaller satellite
of higher density.   They determine that the black hole can exert 
significant tidal forces and disrupt the core of the secondary during
the final pericenter passes.   During previous passes, however,
the secondary will simply be stripped to its tidal radius and the core
will remain intact since its density is greater than that of the 
primary anywhere \citep[see also][for similar results from simulations 
including a black hole in the satellite]{mer01}.   The satellite will 
survive even longer in high
angular momentum encounters (\ie\ those with low eccentricity orbits).
\citet{ken03} find that low angular momentum encounters are more 
efficient in delivering material to the central regions of the primary.   
If the putative minor
merger in Cygnus A is responsible for the triggering of the nuclear
activity, we may be witnessing a more radial encounter after the 
first few pericenter passes or a few $\times 10^7$ yr.    This 
timescale could be consistent with the age estimate (lower limit based on 
synchrotron aging arguments) for the radio source
in Cygnus A of 10$^{6.8}$ yr \citep[][and references therein]{car96}.

One last (unlikely) possibility is that the secondary point source is 
rather reflected light from a cloud with a direct view of the hidden quasar.  
There are several known cases in which a fairly isolated off-nuclear 
cloud has been shown to be dust scattering light from a hidden nucleus,
and even electron-scattering ``mirrors'' are seen off of the nucleus
in some cases.  Particularly instructive is the case of PKS\,2152$-$69,
shown to be highly polarized by \citet{diS88}
and convincingly argued by them to be dust-reflected light from a hidden AGN.
One piece of evidence was the extraordinary ``blueness'' of the spot.
The secondary point source in Cygnus A certainly 
appears to be bluer than the nucleus (see Figs.~\ref{2color} and 
\ref{colorcyga}), but this could be due to different amounts of 
extinction towards each one of these objects, dust emission, and 
the fact that the nuclear $K'$ photometry has a large contribution
from strong emission lines (Paper II).

Several tests can in principle be made of this hypothesis for the secondary
point source in Cygnus A.
High polarization, perpendicular to the radius vector
from the nucleus, would be quite persuasive.  We examined the $HST$ NICMOS
polarization images, originally taken by \citet{tad00}.
However, the marginal spatial resolution, presence of bright
nearby structure, and low signal-to-noise ratio prevented a reliable 
measurement.

A second test is a check for broad emission lines in the putative
scattered nuclear spectrum.  For this test it helps to know in advance
the width of the broad emission lines, and this raises a complication.
The broad \ion{Mg}{2} line width is $\sim7,500$ km sec$^{-1}$, according 
to \citet{ant94}.  These authors
argued that the line was scattered into the line of sight by dust, though
spectropolarimetry was not obtained at that wavelength.
That aperture was, however, placed over
what appears in the optical range as the SE (continuum-emitting) ``nucleus''.
Later \citet{ogl97} found an extremely broad H$\alpha$ (FWHM = 26,000 km
sec$^{-1}$)
line in polarized flux.  Because the ratio of the total flux in this line
to \ion{Mg}{2} is relatively small, and because the width was so much larger,
the simplest picture was that the H$\alpha$ line is scattered by electrons.

For the secondary point source, assuming a line width of 7,500 km s$^{-1}$ 
FWHM, we estimate that any broad Pa$\alpha$ line can have no more than
$\sim$90 \AA\ equivalent width as measured from our spectrum 
(Fig.~\ref{kcomp}).  The limited broad Pa$\alpha$ equivalent width
measurements reported in the literature are much higher, in the range
between 200 and 500 \AA\ \citep[\eg ][]{cou92,kol83}.
This argues against the dust-mirror hypothesis.  Our equivalent width limit
on any emission line with FWHM = 26,000 km s$^{-1}$
like the H$\alpha$ line in polarized flux \citep[][]{ogl97}
is not constraining, but such an
isolated and localized electron-scattering mirror seems unlikely.

Finally, one can consider the reasonableness of the putative scattered
flux given the solid angle subtended by the secondary as seen from the primary.
The secondary has a size $\lesssim$0\farcs05 and a distance from the primary
of 0\farcs4 in projection.  Thus it is likely to subtend no more that
$\sim$1/60 steradians, or $\sim$0.13\% of the total sky as seen from the 
hidden nucleus.
The $\nu$F$_{\nu}$ flux of the secondary point source at 2 $\mu$m is 
($1.5\times10^{14}$ Hz)($\sim$15 $\mu$Jy) = 
$\sim2.2\times10^{-14}$ erg sec$^{-1}$.
Now we need the intrinsic value of $\nu$F$_{\nu}$ at 2 $\mu$m from the primary 
for comparison.
The 2 $\mu$m emission is likely to be highly anisotropic, given its high 
scattering polarization \citep[][]{tad00}.
Instead, we use the nuclear value of $\nu$F$_{\nu}$ at 12 $\mu$m of $\sim$ 120
mJy \citep[][]{why01,rad02} or $\nu$F$_{\nu}  = 9\times 10^{-11}$ 
erg sec$^{-1}$,
which should be more isotropic.  To the present level of accuracy, we can 
assume the same value to apply to the expected 2 $\mu$m $\nu$F$_{\nu}$ 
brightness of the hidden quasar, so the hotspot
has of order $2\times10^{-4}$ times the flux of the hidden nucleus at 2 $\mu$m.
(Here we have assumed unit dust albedo;  this is conservative because we 
are trying to see whether we can rule out sufficient dust reflection for 
the secondary point source.  Also, deviations
from unity \citep[\eg ][Fig.~9]{dra03}
are smaller than geometrical uncertainties in this argument.)
This is consistent with the mirror interpretation, but the argument would
be far more telling if a gross conflict were shown.

\section{Brief Remarks on the Cones\label{cones}}
As mentioned above, apparent edge-brightened ``cones'' are clearly detected 
in our images (Figs.~\ref{knarrow} and \ref{colorcyga}).
\citet{tad99} refer to an ``edge-brightened bicone'' in their 2.0 $\mu$m 
$HST$ images of Cygnus A \citep[see also][for a description of the cones
in the optical]{jac98}.   An analogy is made there to young stellar object 
morphologies and this analogy can be extended to the $\eta$ Carinae nebula 
called the Homunculus.
Similarities exist in both the total-flux and the polarization images, and 
in all
cases these objects can be described, to first order, as bipolar reflection
nebulae.

Our impression is that in some cases the scattered light is seen more 
clearly in
the near-IR than in the optical/UV, because the former wavelengths are less
affected by foreground ``weather'' in the form of dusty gas of modest optical
depth.  This is seen for example in NGC\,1068, in which the far-side scattering
cone is only clear in the near-IR \citep[][]{pac97}.  \citet{ant90}
argued that this is also the case for 3C223.1 and for Cen A.
There are other objects however in which UV dust scattering
in extended merger debris provides the clearest scattered-light picture
of bicones \citep[][]{hur99}.

The edge-brightening in Cygnus A and other objects is likely to result 
in general from a nonuniform distribution of ambient material
\citep[see, \eg ][Sec.~5, for a physical explanation]{bal93}. 
However, a remarkable peculiarity of Cygnus A itself is that, while the
total-flux morphology is symmetric about the cone axis, the polarized flux is
not \citep[][]{tad00}.  The edge of the bicone on which the secondary point
source lies has somewhat less total flux and much less polarization than the
other edge.  The secondary point source may thus provide a clue to the 
different natures of the two bicone edges.  We will discuss this further 
in Paper II (and Kishimoto et al., in preparation, for HST ultraviolet 
imaging polarimetry of this object).

The bicone morphology also begs the question of the role of Cygnus A in the
subject of hidden quasars inside radio galaxies.   Most or even all of the 
highest-luminosity subset of FR\,II narrow-line radio
galaxies have hidden quasars, as shown by polarimetric and other arguments.
By contrast radio galaxies on the lower end of the radio luminosity function 
seem to be heterogeneous with respect to possession of such a (visible or 
hidden) ``thermal'' optical/UV continuum source 
\citep[\eg ][]{why01,ant02,mei01}.
Thus Cygnus A is expected to have a hidden quasar.  This is strongly
suggested by the bicone morphology, and confirmed by the quasar spectrum
observed in scattered light with spectropolarimetry \citep[][]{ogl97}

The exceptional clarity of the apparent cone edges, along with constraints
on the inclination angle, would seem to make our measured opening angle
intelligible in the context of the torus opening angle.  However, such an
inference would be premature because of the $HST$ polarimetric evidence 
cited above \citep[][]{tad00} that the cone edges differ fundamentally 
in character
despite the evident reflection symmetry in the total flux images.  Thus we
postpone any interpretation of the bicone to Paper II.

\section{Summary\label{summary}}

Our high resolution, deep Keck II AO images of Cygnus A show a secondary 
point source in the central regions of Cygnus A.    We have considered
different possibilities in attempting to determine the nature of this object.
The SED and NIR spectrum of the object are inconsistent with those of
foreground Galactic objects.   The most likely explanation is that it is 
indeed an object associated with Cygnus A.   

The absence of ionized gas obviously associated with the object is 
inconsistent with it being a star forming region, and its high luminosity
is orders of magnitude greater than that expected for a globular cluster.
The object has a spectrum and colors that are distinct from those of the 
underlying structure; hence, it does not appear to be a bright compact 
region of the scattering cone which is seen as a discrete object due to 
patchy obscuration.
Finally, the upper limits we place on the presence of broad Pa$\alpha$ in 
the spectrum are inconsistent with a dust mirror interpretation.

We are left with the possibility that the secondary point source in
Cygnus A may be the tidally stripped core of a lower mass galaxy that is
merging with the giant elliptical.   The spectrum of the object is roughly
(though not exclusively) consistent with a that of a dusty compact 
object made of older stars.   The observed luminosity is consistent
with that of a satellite that has lost most of its mass as it is being
accreted, and the observed size ($\lesssim$ 50 pc) is consistent with that 
of an object that has been stripped to its tidal radius. 

The interpretation of a merging low luminosity companion 
is appealing as it would seem to fit in with several observed
properties in Cygnus A.    First, this minor merger could have provided
the means to fuel the black hole, resulting in the nuclear activity
that we observe in Cygnus A.   At least the rough dynamical timescales
seem to be consistent with the estimates for the timescale of the
radio source (both in the order of 10$^7$ yr).   The merger event could
have also triggered the central starburst in Cygnus A that has been
suggested by {\it HST} \citep[][]{fos99,jac98} and ground-based imaging and
spectroscopy \citep[][]{sto94}.   The gas dynamics in the host of Cygnus A
show a peculiar structure with discontinuities, velocity reversals, and
double components akin to the structure found in nearby merger remnants
\citep[][]{sto94}.   The apparent counter-rotating gas in the nuclear
regions is most easily explained as a result of a merger event.
Cygnus A also shows dust lanes which are roughly perpendicular to
the radio jets (see Fig.~\ref{colorcyga}).   A large fraction of
radio galaxies show similarly oriented dust lanes \citep[][]{ver00}
and these are thought to be debris trails of minor mergers
\citep[][]{ken03}.   The jets would then be roughly perpendicular to
the orbit of the encounter that triggered the nuclear activity.   
Cygnus A would fit this scenario, as the putative merging
companion appears to be near the plane of the dust lanes.

However plausible this scenario may seem, it remains highly speculative
at this point.    A deeper AO spectrum showing clearly the presence 
of absorption lines will allow us to determine the precise nature 
and dynamics of this object.   Failing to find absorption lines in
the spectrum, we will need to obtain high angular resolution polarimetry 
to test the hypothesis that the secondary point source is a mirror of
the hidden quasar.   If, however, the secondary point source is confirmed
to be a merging companion, it will be of great interest to model the 
encounter in detail, and Cygnus A will provide, yet again, the testbed
to study important phenomena relevant to powerful active galaxies.

\acknowledgments
We gratefully acknowledge Mark Lacy, Alan Stockton, Luis Ho, 
Makoto Kishimoto, and Bruce Macintosh for helpful
discussions.  We thank Ian McLean and his collaborators for providing
us with their NIRSPEC spectra of L and T dwarfs, and Stephane Charlot
for allowing us to use the GISSEL00 models prior to publication.
We also thank the anonymous referee for helpful suggestions.
Data presented herein were 
obtained at the W.M. Keck Observatory, which is operated as a scientific 
partnership among the California Institute of Technology, the University 
of California and the National Aeronautics and Space Administration. The 
Observatory was made possible by the generous financial support of the
W.M. Keck Foundation.  R.A.'s research is partially supported by NSF
grant AST-0098719.
This work was supported in part by the National Science Foundation 
Science and Technology Center for Adaptive Optics, managed by the 
University of California at Santa Cruz under cooperative agreement 
No. AST-9876783.  This work was supported in part under the auspices 
of the U.S.\ Department of Energy, National Nuclear Security 
Administration by the University of California, Lawrence Livermore 
National Laboratory under contract No. W-7405-Eng-48.

\end{document}